\documentclass[article]{aa}
\def\rounit{{\rm g\,cm^{-3}}}
\def\punit{{\rm dyn\,cm^{-2}}}
\def\eunit{{\rm MeV/nucleon}}
\usepackage{graphicx}
\usepackage{natbib}
\begin{document}
\title{Models of crustal heating in  accreting neutron stars}
\author{P. Haensel
\and  J.L. Zdunik}
 \institute{N. Copernicus Astronomical
Center, Polish Academy of Sciences, Bartycka 18, PL-00-716 Warszawa, Poland
{\em haensel@camk.edu.pl} }
\institute{N. Copernicus Astronomical Center, Polish Academy of Sciences,
Bartycka 18, 00-716 Warszawa, Poland\\
{\tt   haensel@camk.edu.pl, jlz@camk.edu.pl}}
%
%
\abstract{}{Heating associated with non-equilibrium nuclear
reactions in accreting neutron-star crusts is reconsidered,
taking into account suppression of neutrino losses
demonstrated recently by Gupta et al. Two initial
compositions of the nuclear burning ashes, $A_{\rm i}=56$ and
$A_{\rm i}=106$, are considered. Dependence of the integrated
crustal heating on uncertainties plaguing pycnonuclear
reaction models is studied.}{One-component plasma
approximation is used, with compressible liquid-drop
model of Mackie and Baym to describe nuclei.
Evolution of a crust shell is followed
from $10^8~{\rm g~cm^{-3}}$ to $10^{13.6}~{\rm g~cm^{-3}}$.}
{The integrated heating in the outer crust agrees nicely with results
of self-considtent multicomponent plasma simulations of
Gupta et al.; their results fall between our curves obtained
for  $A_{\rm i}=56$ and
$A_{\rm i}=106$. Total crustal heat per one accreted nucleon
ranges between $1.5~\eunit$ to $1.9~\eunit$ for $A_{\rm i}=106$ and
$A_{\rm i}=56$, respectively. The value of $Q_{\rm tot}$
depends weakly on the presence of pycnonuclear reactions at
$10^{12}-10^{13}~\rounit$. Remarkable  insensitivity of $Q_{\rm tot}$
on the details of the distribution of nuclear processes in
accreted crust is explained.}{}

\titlerunning{Crustal heating in accreting neutron stars}
\maketitle
\section{Introduction}
\label{sect:introduction}
Neutron star crust that is not in full thermodynamic equilibrium constitutes
a reservoir of energy, which can then be released during star's evolution.
Formation and structure of non-equilibrium neutron star crust was considered
by many authors  (\citealt{Vartanyan1976,BisnoChech1979,
Sato1979,HZ1990,HZ2003,Gupta2007}).
Such a state of the crust can be produced as a result of accretion of
the matter onto a neutron star in a close low-mass X-ray binary, where
the original crust built of a catalyzed matter could actually
be  replaced by a crust with composition strongly
deviating from the nuclear equilibrium one.

Heating due to non-equilibrium nuclear
processes taking place in the outer and inner crust of an accreting neutron star
(deep crustal heating) was calculated by \cite{HZ1990} who used
a simple model of  one-component plasma, and assumed that the outer layers of the
matter produced  in the X-ray bursts
consisted of  pure $^{56}{\rm Fe}$.   Another simplification used by
\cite{HZ1990} consisted in assuming  the ground-state - ground-state
nuclear transitions due to the electron captures. Consequently, they maximized
neutrino losses. Their calculated total deep crustal heating,
produced mainly in the inner crust, was $Q_{\rm tot}
\sim 1.4~{\rm MeV}$ per one accreted
nucleon. \cite{HZ2003} recalculated deep crustal heating for different initial
composition of the outer layers, and obtained similar values of
$Q_{\rm tot}=1.2-1.4~$MeV/nucleon.  Recently, heating of the outer
 crust of an accreting neutron star was
studied by \cite{Gupta2007} who went beyond  a simple model
of \cite{HZ1990} and \cite{HZ2003}. Namely, \cite{Gupta2007}
considered a multicomponent plasma, a reaction network
of many nuclides,  and included the contribution from
the nuclear excited states. They found
that electron captures in the outer crust lead mostly to
the excited states of the
daughter nuclei, which then deexcite heating the matter.
Consequently, they found
that the neutrino losses in the outer crust were negligible, which
strogly increased the outer crust heating,
compared to \cite{HZ2003} (by a factor of four).
 However, outer crust contributes only
a small fraction of the $Q_{\rm tot}$, and neutrino losses in the inner crust,
where the bulk of $Q_{\rm tot}$ is produces are small,
 so that the original value
of \cite{HZ1990}, $Q_{\rm tot}\sim 1.4~$MeV/nucleon,
remains quite a reasonable estimate
(see Sect.\ \ref{sect:heating} of the present paper).

The phenomenon of deep crustal heating appears  to be relevant for the understanding
of the thermal radiation observed in the soft X-ray transients (SXTs) in
quiescence, when the accretion from a disk formed of
plasma  flowing from the low-mass companion
star is switched off or strongly suppressed.
Typically, the quiescent emission is much higher than the expected one for an old
 neutron star. It has been suggested that this is due to the fact that the
interiors of neutron stars in SXTs are heated-up,  during relatively short periods of
accretion and bursting, by the non-equilibrium processes associated with nuclear
reactions taking place in the deep layers of the crust (Brown et al. 1998). The
deep crustal heating, combined with appropriate models of neutron-star
atmosphere and interior, is   used to explain observations of SXTs in quiescence.
The luminosity in quiescence depends on the structure of neutron-star core, and
particularly on the rate of neutrino cooling. This   opened a new possibility
of exploring the internal structure and equation of state of neutron stars via
confrontation of theoretical models with observations of quiescent SXTs
(see \citealt{ColpiGeppert2001,Rutledge2002,YakLH2003,YakLP2004,LevH2007} ).

Up to now, the crustal heating used in modeling SXTs was described
using the model of  Haensel \& Zdunik (1990)(hereafter referred to as HZ90), updated
and generalized by  Haensel \& Zdunik (2003)(hereafter referred to as HZ03).
In these models, the heat was produced during the active (accretion) episodes,
when the outer layer  of neutron star
was   sinking in the neutron star interior under the weight of
accreted matter. The very outer layer was assumed to be composed of the ashes of the
X-ray bursts in the active epoch.  For simplicity, those ashes  were assumed to be
a one-component  plasma  ($^{56}{\rm Fe}$ in HZ90 and $^{56}{\rm Fe}$ and
$^{106}{\rm Pd}$ in HZ03).  Under an increasing pressure, the composition of a
sinking matter element was  changing in a sequence of nuclear reactions:
electron capture, neutron emission and absorption, and finally,
at densities exceeding
$10^{12}~{\rm g~cm^{-3}}$, also pycnonuclear fusion.
As the nuclear processes were   proceeding off-equilibrium, they
were accompanied by the heat deposition in the crustal matter.
 The crustal  heating was mostly supplied by the
pycnonuclear fusion processes in the inner crust at $\rho=10^{12}-10^{13}~{\rm
g~cm^{-3}}$. This seemingly crucial r{\^o}le of pycnonuclear reactions is
embarassing, because their rates are plagued by huge uncertainties.
As shown by \cite{YakGW2006}, the uncertainty
in the calculated rate of pycnonuclear fusion of two $^{34}{\rm Ne}$ nuclei,
first pycnonuclear fusion in the inner crust as predicted by the HZ90 model,
can be as large as seven (!) orders of magnitude.
Therefore, there is a basic uncertainty about which
pycnonuclear fusions do occur
and at what densities. Fortunately, as we show
in the present paper, this uncertainty
does not affect significantly the value of $Q_{\rm tot}$.
However, it implies an ignorance
 concerning the distribution of heat sources in the inner crust. Possible
observational constraints on the distribution of heat sources in the inner
crust are discussed in Sect.\ \ref{sect:conclusion}.

In the present paper  we redo the calculations of the crustal heating taking into account
uncertainties in the pycnonuclear reaction rates as well as the suppression of
the neutrino losses. We show that the uncertainties in the pycnonuclear reaction
rates do not significantly affect the total heat release in the crust, while influencing
of course the radial distribution of heat sources. We also show that the
effect of excited states of daughter nuclei, while very important in the outer
crust, does not lead to a significant increase of total crustal heating.
Calculations are performed assuming  two different initial
nuclides  produced in the X-ray bursts.

The plan of the paper is as follows. In Sect.\ \ref{sect:processes} we
briefly remind the scenario of the deep crustal heating, nuclear model
used in simulations, and describe the relevant physical processes
acting in the crust of an accreting neutron star. Results of
selected simulations of the
nuclear evolution of a matter element compressed from $10^8~\rounit$
to $10^{13.6}~\rounit$,
 are presented in the form of figures in Sect.\ \ref{sect:networks}
 and  tables in the Appendix. Total crustal heating is calculated
 in Sect.\ \ref{sect:heating}.
In Sect.\ \ref{sect:constQ} we give an explanation  of the weak model dependence
of the total crustal heating (per one accreted nucleon). Our conclusions
are presented in Sect.\ \ref{sect:conclusion}, where we
also suggest an observational testing of the actual radial
distribution of heat sources,
which could be helpful in putting constraints on the deep crustal
heating models.

\section{Non-equilibrium nuclear processes}
\label{sect:processes}
\subsection{Outer crust}
\label{sect:processes-outer_crust}
\begin{figure}
\resizebox{\hsize}{!}{\includegraphics[angle=0]{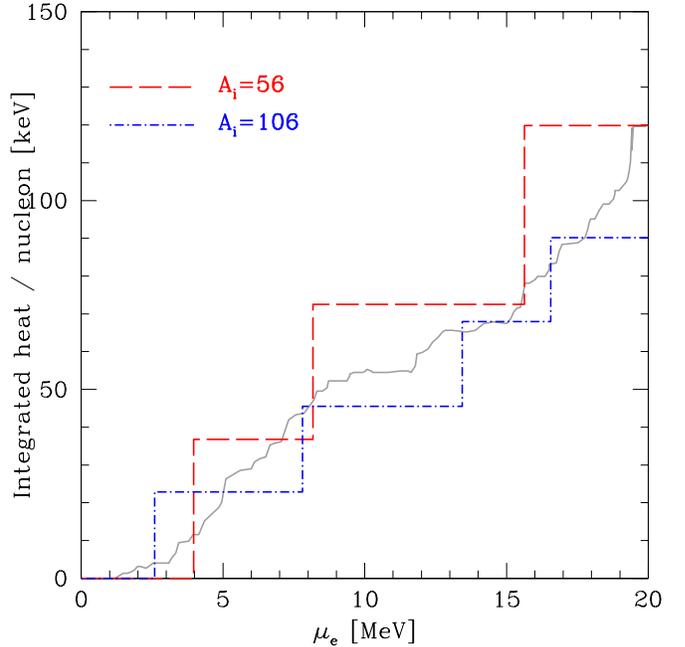}}
\caption{(Color online) Integrated  crustal heating per one accreted nucleon
versus electron Fermi energy $\mu_e$. The step-like curves
were obtained using the HZ* model of the present paper, for
two choices of an initial nuclide. The continuous curve was obtained by
\cite{Gupta2007}. }
\label{fig:compareQ}
\end{figure}
In what follows we briefly describe the nuclear evolution scenario
of HZ90 and HZ03, with correction implied by the results of
\cite{Gupta2007}. Under conditions prevailing in accreting neutron-star
crust at $\rho>10^8~{\rm g~cm^{-3}}$ matter is strongly
degenerate, and is relatively cold ($T < 10^8~{\rm K}$), so that
thermonuclear processes involving charged particles
can be assumed to be  blocked by the Coulomb barrier.
Consequently,  the densities lower than the threshold for the pycnonuclear
fusion (which is very uncertain, see \citealt{YakGW2006},
$\rho_{\rm pyc}\sim 10^{12}-10^{13}~{\rm g~cm^{-3}}$), the number of
nuclei in an element of matter does not change during the
compression resulting from the increasing weight of accreted
matter. Let us remind, that we assume that only one nuclear species
$(A,Z)$ is present at each pressure (one component plasma).
Due to the nucleon pairing,
stable nuclei in dense matter  have even $N=A-Z$ and $Z$
(even-even nuclides). In the outer crust, in which free neutrons are
absent, the electron captures which proceed in two steps,
\begin{eqnarray}
(A,Z)+e^-&\longrightarrow & (A,Z-1)+\nu_e~, \cr\cr (A,Z-1)+e^-&\longrightarrow &
(A,Z-2)+\nu_e + Q_j~. \label{eq:e.cap}
\end{eqnarray}
lead to a systematic decrease of $Z$ (and increase of $N=A-Z$)
with increasing density.  The first capture in Eq.\
(\ref{eq:e.cap}) proceeds in a quasi-equilibrium manner, with
a negligible energy release. It produces an odd-odd nucleus,
which is strongly unstable in dense medium and captures a
second electron in an non-equilibrium manner, with energy
release $Q_j$, where $j$ is the label of the non-equilibrium
process.
\subsection{Outer crust: comparison of HZ* model
with \cite{Gupta2007}}
\label{sect:modifiedHZ}
In the  original model of HZ90 electron captures were assumed
to proceed from the ground
state of the initial nucleus to the ground state of the
daughter nucleus (GS-GS), and therefore most of the
energy release was taken away by neutrinos (from $3/4$
to $5/6$ of $Q_j$,  see \citealt{HZ2003} for the discussion
of this point). Very recently,
an extensive, multicomponent self-consistent calculation of the nuclear
evolution of an accreted matter element in the crust of a
bursting neutron star was carried out by \cite{Gupta2007}.
As they have shown, electron captures to excited states (GS-ES) and
subsequent de-excitation strongly reduce neutrino
losses, so that  nearly all released energy heats the
crust matter. We therefore modified the original HZ model
by neglecting neutrino losses accompanying electron
captures: this model will be denoted as HZ*. Consider a
version HZ* model, corresponding
to specific choice of the the initial mass number $A_{\rm i}$,
and  denoted by a label $(\alpha)$. The integrated heat deposited
in the crust in the layer with bottom density $\rho$ can then
be obtained as
\begin{equation}
Q^{(\alpha)}(\rho)=
\sum_{j(\rho_j<\rho)}Q^{(\alpha)}_j~.
\label{eq:Qcumul}
\end{equation}
In Fig.\ \ref{fig:compareQ} we compare $Q(\mu_e)$, obtained
for two HZ* models, with results of \cite{Gupta2007}. We
replaced $\rho$ variable  by the electron Fermi energy $\mu_e$
to facilitate the comparison. The curve of
\cite{Gupta2007} lies between two extreme  HZ* models; it has
a very large number of small jumps.  These two features
reflect the multicomponent structure of crustal matter and
large network of nuclear reactions in
\cite{Gupta2007}, with $50\la A_{\rm i} \la 110$.
Generally, viewing the simplicity of the HZ* model, its
agreement with multicomponent, self consistent calculations of
\cite{Gupta2007}, is very good.
\subsection{Inner crust}
\label{sect:processes-inner_crust}
\begin{figure}
\resizebox{\hsize}{!}{\includegraphics[angle=0]{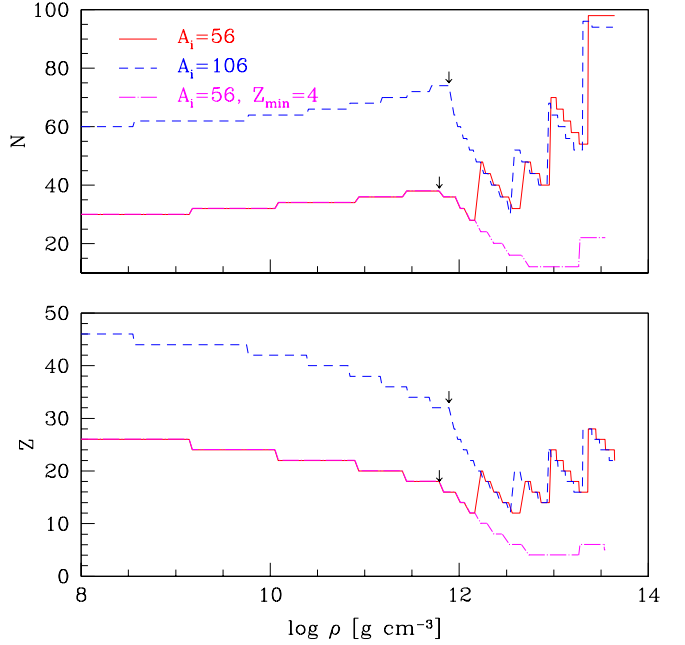}}
\caption{
(Color online) $Z$ and $N$ of nuclei, versus matter density in an
accreting neutron-star crust, for different  models of dense matter.
Solid line: $A_{\rm i}=56$; dashed line: $A_{\rm i}=106$.
Dash-dot line: continuation of $A_{\rm i}=56$ evolution but
with pycnonuclear fusion blocked
until $Z=Z_{\rm min}=4$. Each change of $N$ and $Z$, which takes
place at a constant pressure, is accompanied by a jump in density
(see HZ90 for detailed discussion of this point). Small steep
segments connect the top and the bottom density of thin reaction
shell. Arrows indicate positions of the neutron drip point.
}
 \label{fig:NZ_rho_pyc}
\end{figure}
 Above the neutron-drip point
 ($\rho>\rho_{\rm ND}$), electron captures trigger neutron emissions,
\begin{eqnarray}
(A,Z)+e^-&\longrightarrow & (A,Z-1)+\nu_e~, \cr\cr (A,Z-1)+e^-&\longrightarrow &
(A-{\rm k},Z-2)+{\rm k}\; n + \nu_e + Q_j~.
 \label{eq:e.cap.n}
\end{eqnarray}
 Due to the electron captures, the
value of $Z$ decreases with increasing density. In consequence, the Coulomb barrier
prohibiting the nucleus-nucleus reaction lowers. This effect, combined with the
decrease of the mean distance between the neighboring nuclei, and a simultaneous increase
of energy of the quantum zero-point vibrations around the nuclear lattice sites, opens
a possibility of the pycnonuclear  reactions. The pycnonuclear fusion timescale
$\tau_{\rm pyc}$ is a very sensitive function of $Z$. The chain of the reactions
(\ref{eq:e.cap.n}) leads to an abrupt decrease of
$\tau_{\rm pyc}$ typically by 7 to 10 orders of magnitude.
Pycnonuclear fusion switches-on as
soon as $\tau_{\rm pyc}$ is smaller than the time of the
 travel  of a piece of matter (due to the accretion) through the considered
shell of mass $M_{\rm shell}(N,Z)$,
$\tau_{\rm acc}\equiv M_{\rm shell}/\dot{M}$. The masses of the shells are
of the order of $10^{-5}~ {\rm
M}_\odot$. As a result, in the inner crust the
chain of reactions (\ref{eq:e.cap.n}) in several cases is
followed by the pycnonuclear reaction on a timescale
much shorter than $\tau_{\rm acc}$.
Denoting $Z'=Z-2$, we have then
\begin{eqnarray}
&~&~~~~~~~~(A,Z')+(A,Z')\longrightarrow  (2A,2Z')+Q_{j,1}~,\cr\cr &~&(2A,2Z')\longrightarrow
(2A-{\rm k'}, 2Z')+{\rm k'}\;n + Q_{j,2}~,\cr\cr &~&\ldots~~~~~~\ldots~~~~~~\ldots~~~~~~\ldots
+Q_{j,3}~,
\label{eq:pyc.scheme}
\end{eqnarray}
where ``$\dots$'' correspond to some not specified chain of the
electron captures accompanied by neutron emissions. The total heat
deposition in matter, resulting from a chain of reactions
involving a pycnonuclear fusion, is $Q_j=Q_{j,1}+Q_{j,2}+Q_{j,3}$.
In contrast to HZ03, and in accordance with \cite{Gupta2007},
we neglect the neutrino losses accompanying the non equilibrium
electron captures.

Our model of atomic nuclei is described in detail
in HZ90. Using our nuclear-evolution code, we
evolved an element of matter  consisting initially of nuclei $(A_{\rm i},Z_{\rm i})$
immersed in an electron gas, from $\rho_{\rm i}= 10^8~{\rm g~cm^{-3}}$ to
$\rho=\rho_{\rm f}>5\times 10^{13}~{\rm g~cm^{-3}}$. Our results for the composition and
crustal heating are presented in the next section and in the Appendix.
\section{Results of simulations: reactions,  heating, compositions}
\label{sect:networks}
The compositions of accreted neutron star crusts are shown
in Fig. \ref{fig:NZ_rho_pyc} and in tables in the Appendix.
These results describe  crusts built of accreted and processed matter up to the
density $5\times 10^{13}~{\rm g~cm^{-3}}$. At a constant accretion rate $\dot{M}=\dot{M}_{-9}\times 10^{-9}~{\rm
M}_\odot/{\rm yr}$ this will take $\sim 10^6~{\rm yr}/\dot{M}_{-9}$. During
that time, a shell of X-ray burst ashes will be compressed from
$\sim 10^8~{\rm g~cm^{-3}}$ to $\sim 10^{13}~{\rm g~cm^{-3}}$.

Two different compositions of X-ray bursts ashes
at $\la 10^8~ {\rm g~cm^{-3}}$ were assumed. In the first case,
$A_{\rm i}=56$, $Z_{\rm i}=26$,  like in HZ90.
In the second scenario we followed
HZ03, with $A_{\rm i}=106$,  to imitate nuclear ashes
obtained by \cite{Schatz2001}. The
value of $Z_{\rm i}=46$ stems then from the condition of beta equilibrium at
$\rho=10^{8}~{\rm g~cm^{-3}}$.  The density dependence of $Z$
and $N=A-Z$ within the accreted crust is shown in Fig.\
\ref{fig:NZ_rho_pyc}. After pycnonuclear fusion region has
been reached, both curves converged (as explained in HZ03,
this results from $A_{\rm i}$ and $Z_{\rm i}$ in two scenarios
differing by a factor of about two). Suppressing pycnonuclear
fusion in the $A_{\rm i}=56$ case until $Z=Z_{\rm min}=4$
yields the lowest curve.

\begin{figure}
\resizebox{\hsize}{!}{\includegraphics[angle=0]{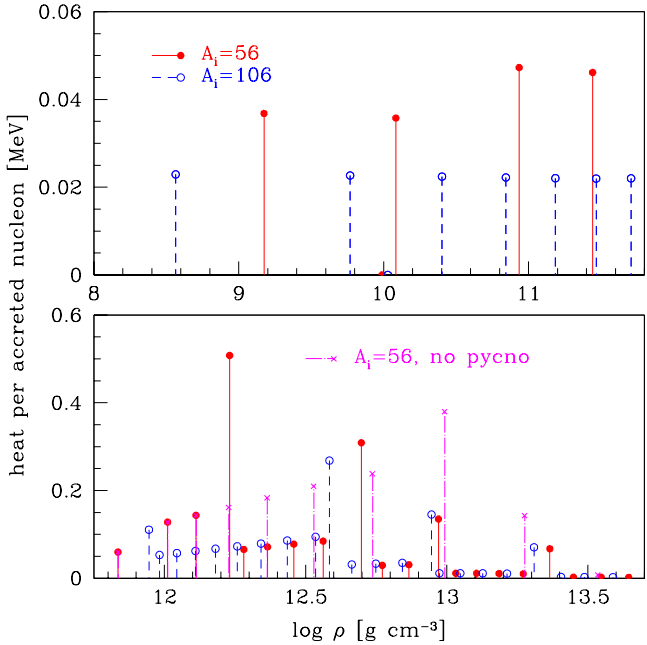}}
\caption{(Color online) Heat sources in the outer (upper panel) and inner (crust)
for three HZ* models. Vertical lines, positioned at the density at
the bottom of the reaction shell,  represent the heat
per one accreted nucleon. Labels as in Table 2.
 }
\label{fig:H_sources}
\end{figure}
In Fig.\ \ref{fig:H_sources} we show the heat deposited in the matter, per one accreted
nucleon, in the thin shells in which non-equilibrium nuclear processes are taking
place. Actually, reactions proceed at a constant pressure, and there is a density jump
within a thin  ``reaction shell''. The vertical lines whose height gives  the heat
deposited in matter are drawn at the density at the  bottom  of the reaction shell.

One notices a specific dependence of the number of heat sources and the heating
power of a single source on assumed $A_{\rm i}$.
 In the case of $A_{\rm i}=56$ the number of sources is
 smaller, and their heat-per-nucleon values $Q_j$ are
 significantly larger,  than for $A_{\rm i}=106$.
\section{Integrated  crustal heating}
\label{sect:heating}
\begin{figure}
\resizebox{\hsize}{!}{\includegraphics[angle=0]{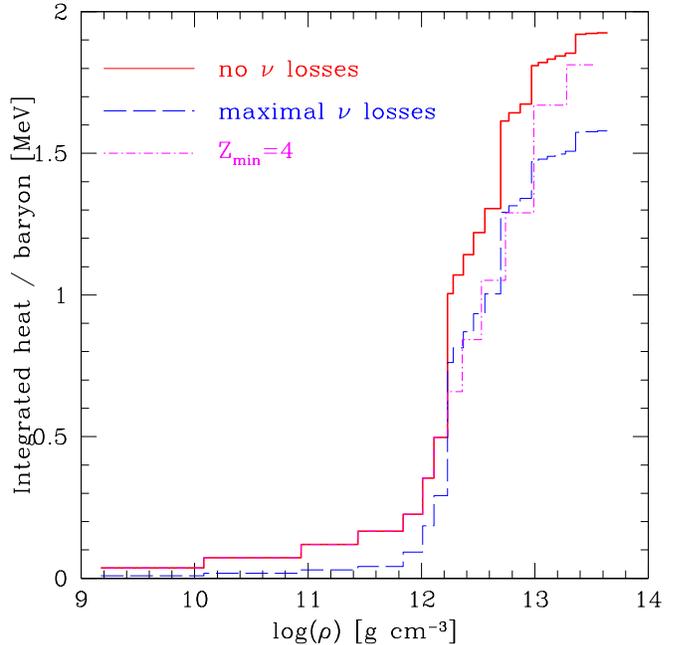}}
\caption{(Color online) Integrated heat released in the crust,
$Q(\rho)$ (per one accreted nucleon) versus $\rho$, assuming
initial ashes of pure $^{56}{\rm Fe}$. Solid line: HZ* model
of the present paper, with suppressed neutrino losses. Long
dashes: GS-GS transitions in electron captures, with maximal
neutrino losses. Dash-dot line: No neutrino losses, with
pycnonuclear fusion blocked until $Z=Z_{\rm min}=4$.
 }
\label{fig:ecumul}
\end{figure}
The quantity $Q^{(\alpha)}(\rho)$, for three specific models
of compressional evolution,  is plotted in Fig.\ \ref{fig:ecumul}.
In the all three cases, we set $A_{\rm i}=56$ and $Z_{\rm i}=26$.
For the first model, we neglect neutrino losses; its cumulated heat $Q^{(1)}$
is always the highest. The second   model is  used to visualize the important
of excited states of the daughter nuclei in the electron captures. For this model
we assumed  that the nuclear transitions associated with electron captures
are of the GS-GS type, which maximizes the neutrino losses. While the
effect is dramatic for  $\rho\la 10^{12}~\rounit$, it implies only
a 20\% underestimate of $Q$ above $10^{12.5}~\rounit$. But the most interesting
is perhaps the effect of literally switching off of the pycnonuclear reactions,
assumed in the third scenario. This was done by assuming that the pycnonuclear
fusion   is blocked till the nuclear charge goes down to $Z_{\rm min}=4$.
 And yet, for $\rho_{\rm b}>10^{13}~\rounit$,  $Q^{(3)}$ is very similar
to that obtained in the first scenario,  which was most advantageous
as far as the crust heating was concerned.
Namely, a missing pycnonuclear heating at $\rho\sim 10^{12}~\rounit$ is
efficiently compensated by the electron captures accompanied by neutron emission
within the density decade $10^{12}-10^{13}~\rounit$.
The values of $Q$ saturate above $10^{13.6}~\rounit$, where
 80\% of nucleons are in
neutron gas phase. All in all, for three scenarios with $A_{\rm i}=56$,
the total deep crustal heat release is $(1.6-1.9)~{\rm MeV}$/nucleon.
For $A_{\rm i}=106$,  numbers are shifted downward by about 0.4 MeV/nucleon.
The summary of our results for the total heat release is given in
Table\ \ref{tab:Qtot}.

\begin{table*}[t]
\begin{center}
\caption{Total crustal heating $Q_{\rm tot}$ for $A_{\rm i}=56$ and
$A_{\rm i}=106$. First line:  HZ* model of the present paper.
Second line: old HZ03 model with maximal neutrino losses. Third and
 fourth lines: results obtained when neutrino losses are
 suppressed  and  pycnonuclear fusion is blocked down to $Z_{\rm min}=6$
 and $Z_{\rm min}=4$, respectively.  }
\begin{tabular}{ccc|ccc}
\hline
 & $A_{\rm i}=56 $ &  &   &$A_{\rm i}=106 $ &    \\
 \hline
   pycno  & $\nu$ losses &  $Q_{\rm tot}$ &
 pycno  & $\nu$ losses &  $Q_{\rm tot}$ \\
\hline
no blocking  &  none  & 1.93 MeV &
no blocking  &  none  &  1.48 MeV
\\
no blocking  &  maximal  & 1.58 MeV &
no blocking  &  maximal  &  1.16 MeV
\\
 $Z_{\rm min}=6$  &    none & 1.93 MeV &
 $Z_{\rm min}=6$  &  none  & 1.44 MeV
\\
 $Z_{\rm min}=4$  &  none  & 1.85 MeV &
 $Z_{\rm min}=4$  &  none  & 1.35 MeV
 \\
\hline
\end{tabular}
\label{tab:Qtot}
\end{center}
\end{table*}
\section{Constancy of the total heat release}
\label{sect:constQ}
\begin{figure}
\resizebox{\hsize}{!}{\includegraphics[angle=0]{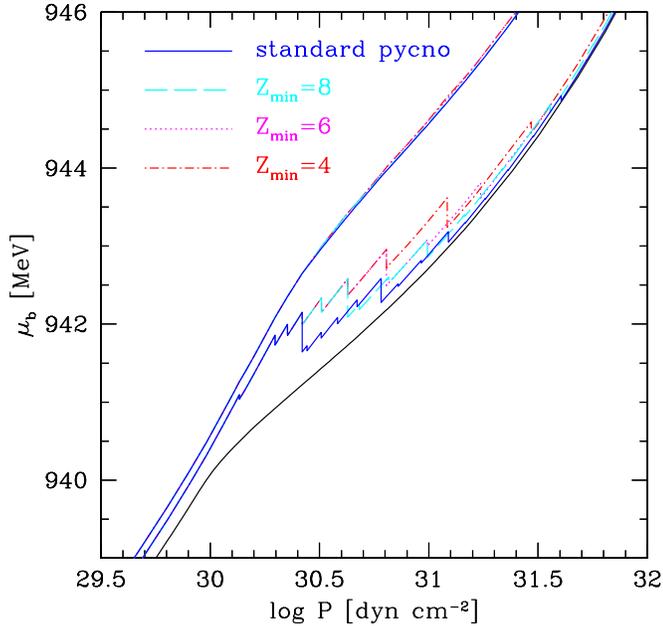}}
\caption{(Color online) Baryon chemical potential $\mu_{\rm b}(P)$ for
different versions of blocking of pycnonuclear fusion.
"Standard pycno" corresponds to HZ* model of the present
paper, with $A_{\rm i}=56$ . Three other curves
correspond to pycnonuclear fusion suppressed until to $Z=Z_{\rm
min}=8,6,4$, respectively. The upper continous curves, which
nearly coincide,  are defined as
$\overline{\mu}^{(\alpha)}_{\rm b}(P)
\equiv {\mu}_{\rm b}^{(\alpha)}(P)
+ \sum_{j
(P_j<P)}Q^{(\alpha)}_j$. Dependence
of $\overline{\mu}^{(\alpha)}_{\rm b}(P)$ on $(\alpha)$ is
negligible, so that
$\overline{\mu}^{(\alpha)}_{\rm b}(P)\approx \overline{\mu}_{\rm b}(P)$.
The lowest smooth solid curve - cold
catalyzed matter. }
 \label{fig:mup}
\end{figure}
Remarkably weak dependence of the total heat release in the crust,
$Q_{\rm tot}$, on the nuclear history of an element of matter
 undergoing compression from $\sim 10^{8}~\rounit$ to $\sim 10^{13.6}~\rounit$,
 deserves an explanation. Let us therefore study the most relevant thermodynamic
quantity, the Gibbs free energy per nucleon, which
is the baryon chemical potential, $\mu_{\rm b}$.
In the  $T=0$ approximation,  $\mu_{\rm b}(P)=[{\cal E}(P)+P]/n_{\rm b}(P)$.
Minimizing  $\mu_{\rm b}(P)$, at a fixed $P$,  with respect to the independent
thermodynamical variables  ($A,Z$, mean free neutron density $\overline{n}_n$,
mean baryon density $n_{\rm b}$, size of the Wigner-Seitz cell, etc.),
under the constraint of electric charge neutrality,
 $\overline{n}_p=\overline{n}_e$, we get the the ground state of the
 crust a given $P$.  This state is
called "cold catalyzed" matter, and its
baryon chemical $\mu^{(0)}_{\rm b}(P)$
is represented   in Fig.\ \ref{fig:mup} by a solid black line.
All other $\mu^{(\alpha)}_{\rm b}(P)$ curves, showed in
Fig.\ \ref{fig:mup}, display discountinuous drops
due to non-equilibrium reactions included in a given evolutionary
model $(\alpha)$, and lie above the solid black one. This  visualizes the
fact that non-catalyzed matter is a reservoir of energy, released in
non-equilibrium processes that push down matter closer to the absolute
ground state. In spite of dramatic differences between different
$\mu^{(\alpha)}_{\rm b}(P)$ in the region where the bulk of non-equilibrium
reactions and heating occur,  $P=(10^{30}-10^{31.5})~\punit$,
these functions tend to the ground
state one for $P\ga 10^{32}~\punit$. The general structure of
different $\mu^{(\alpha)}_{\rm b}(P)$ is similar. At same $P$, their continuous
 segments have nearly the same slope. What differs $\mu^{(\alpha)}_{\rm b}(P)$
   are discontinuous drops, by $Q^{(\alpha)}_j$,
at reaction thresholds $P^{(\alpha)}_j$. The functions
$\mu^{(\alpha)}_{\rm b}(P)$ can therefore be expressed as
\begin{equation}
\mu^{(\alpha)}_{\rm b}(P)\approx
\overline{\mu}_{\rm b}(P)
- \sum_{j (P_j<P)}Q^{(\alpha)}_j,
 \label{eq:decomp-mu}
\end{equation}
where $\overline{\mu}_{\rm b}(P)$ is a smooth function of $P$,
independent of $(\alpha)$ (for explanation, see Fig.\ \ref{fig:mup}
and its caption).
For $P>10^{33.5}~\punit$ the values
of $Q_j$ are negligibly small, and   all
$\mu^{(\alpha)}_{\rm b}(P)$ become quite close to the ground state line.
This implies, that the sum $\sum_jQ^{(\alpha)}_j$ has to be,
to a rather good precision,  {\it independent} of ${(\alpha)}$.
\section{Discussion and conclusion}
\label{sect:conclusion}
In the present paper we reconsidered a model of deep crustal
heating, formulated originally in (\citealt{HZ1990,HZ2003}).
 Following \cite{Gupta2007}, we suppressed neutrino losses
 associated with electron captures.
In this way we got the HZ* model of crustal heating that
despite its simplicity (one component plasma) agrees nicely
with results of self-consistent multi-component calculations of
\cite{Gupta2007}. Using  the HZ* model we obtained the  total crustal
heating $Q_{\rm tot}=1.5~$MeV and 1.9~MeV/nucleon, for the initial
ashes consisting of $^{106}{\rm Pd}$ and $^{56}{\rm Fe}$, respectively.
We studied the dependence of crustal heating on the location of the pycnonuclear fusion
processes within the crust. It turned out that the total
crustal heating (per one accreted nucleon) is quite
insensitive to the depth at which pycnonuclear fusion occurs,
with $Q_{\rm tot}$ varying at most by 0.2 MeV,
and we presented an explanation of this feature.

Maximal neutrino losses, implied by the assumption of the
ground-state - ground-state electron captures in \cite{HZ1990}
and \cite{HZ2003}, led to a severe underestimate of the
heating of the outer crust (\citealt{Gupta2007}). However,
the outer crust gives a rather small contribution to the total
crustal heating, $Q_{\rm tot}$, and the underestimate of
$Q_{\rm tot}$ is less than  25\%. Composition of the initial
ashes of the X-ray bursts can be more important for the value of
$Q_{\rm tot}$, as shown already in (\citealt{HZ2003}).

The insensitivity of $Q_{\rm tot}$ to the very uncertain rates
of pycnoclear fusion means that this tremendous uncertainty,
unlikely to be removed in spite of future theoretical efforts,
does not affect the thermally equilibrated quiescent state of
the SXTs. This may seem as good news. On the other hand, this
means that the studies of SXTs in quiescence will not improve our
knowledge of pycnonuclear fusion in dense plasma: this is bad
news.

Fortunately, the situation changes with the
access to observations of the thermal relaxation
in SXTs after the accretion episode. This
phenomenon cannot be observed in standard SXTs that remain in
accreting state for days - weeks. Thermal relaxation can be
observed only in so called persistent SXTs, characterized by
accretion states lasting for years - decades. Such thermal
relaxation, called initial cooling, has been observed in
KS 1731-260 and MXB 1659-29 (\citealt{Cackett2006}). Let
us focus on  KS 1731-260.  After  12.5 years of
accretion and associated crustal heating, the crust of KS
1731-260 has become significantly hotter than the neutron star
core. After accretion stopped, the heat cumulated in the
crust diffused over the star, and the stellar surface cooled.
The  cooling curve of KS 1731-260 toward the quiescent state  has been
obtained by \cite{Cackett2006}. This curve depends on
the crust thermal conductivity, crust thickness, distribution
of crustal heating sources, and on the neutrino cooling of
neutron star core (\citealt{Rutledge2002,Cackett2006}).

There are two complementary aspects of observations of
cooling of persistent SXTs. On the one hand, we need a theory to
understand this phenomenon. On the other hand, observational
data yield constraints on theoretical models.
Very recently, simulations of thermal relaxation of KS
1731-260 were performed along these lines  by
\cite{Shternin2007}. Increasing in number and precisions,
observations of cooling curves in persistent SXTs will hopefully
be a promising testing ground for the theories  of deep
crustal heating, including pycnonuclear fusion, and other physical processes in
neutron stars.
\begin{acknowledgements}
We are very grateful to D.G. Yakovlev for a critical reading
of the manuscript and for helpful remarks.
This work was supported in part by the KBN grant 5 P03D 020 20
and the MNiSW grant N20300632/0450.
\end{acknowledgements}

\begin{appendix}
\section{Tables}
\newcommand {\st}{\rightarrow}
\begin{table*}[t]
\caption{
Non-equilibrium processes in the crust of an accreting neutron stars assuming
that the X-ray ashes consist of pure $^{106}{\rm Pd}$.
$P_j$ and $\rho_j$ are pressure and density at which the reaction
takes place.  $\Delta \rho/\rho_j$ is relative density jump
connected with reaction, $Q_j$ is the heat deposited in the
matter. $X_n$ is the fraction of free neutrons
among  nucleons, and $\mu_e$ is the electron chemical potential,
both in the layer just above the reaction surface.
}
\label{tab:crust106}
\begin{center}
\begin{tabular}{llllrrr}
\hline\hline $P_j$ & $\rho_j$ & reactions & $X_n$  & $\Delta\rho/\rho_j$ & $\mu_{\rm e}$ & $Q_j$\\
  (dyn~cm$^{-2}$)  &   (g~cm$^{-3}$)& & &\% & (MeV) & (keV)\\
\hline
 $  9.235\times 10^{25}$ &  $ 3.517\times 10^{08}$ &   $^{106}{\rm Pd}\st ^{106}{\rm Ru}-2e^-+2\nu_e $ & $     0      $  &$  4.4  $&$  2.29 $&$    22.9$ \\
 $  3.603\times 10^{27}$ &  $ 5.621\times 10^{09}$ &   $^{106}{\rm Ru}\st ^{106}{\rm Mo}-2e^-+2\nu_e $ & $     0      $  &$  4.6  $&$  6.34 $&$    22.7$  \\
 $  2.372\times 10^{28}$ &  $ 2.413\times 10^{10}$ &   $^{106}{\rm Mo}\st ^{106}{\rm Zr}-2e^-+2\nu_e $ & $     0      $  &$  4.9  $&$ 10.43 $&$    22.4$  \\
 $  8.581\times 10^{28}$ &  $ 6.639\times 10^{10}$ &   $^{106}{\rm Zr}\st ^{106}{\rm Sr}-2e^-+2\nu_e $ & $     0      $  &$  5.1  $&$ 14.56 $&$    22.2$  \\
 $  2.283\times 10^{29}$ &  $ 1.455\times 10^{11}$ &   $^{106}{\rm Sr}\st ^{106}{\rm Kr}-2e^-+2\nu_e $ & $     0      $  &$  5.4  $&$ 18.73 $&$    22.1$  \\
 $  5.025\times 10^{29}$ &  $ 2.774\times 10^{11}$ &   $^{106}{\rm Kr}\st ^{106}{\rm Se}-2e^-+2\nu_e $ & $     0      $  &$  5.7  $&$ 22.91 $&$    22.0$  \\
 $  9.713\times 10^{29}$ &  $ 4.811\times 10^{11}$ &   $^{106}{\rm Se}\st ^{106}{\rm Ge}-2e^-+2\nu_e $ & $     0     $  &$  6.1   $&$ 27.08 $&$     22.0$  \\
 \hline
 $  1.703\times 10^{30}$ &  $ 7.785\times 10^{11}$ &   $^{106}{\rm Ge}\st ^{92}{\rm Ni}+14n-4e^-+4\nu_e $ & $  0.13 $  &$ 13.2  $&$ 31.22 $&$  110.8$  \\
 $  1.748\times 10^{30}$ &  $ 8.989\times 10^{11}$ &   $ ^{92}{\rm Ni}\st ^{86}{\rm Fe}+ 6n-2e^-+2\nu_e $ & $  0.19 $  &$  6.9  $&$ 31.33  $&$   53.2$  \\
 $  1.924\times 10^{30}$ &  $ 1.032\times 10^{12}$ &   $ ^{86}{\rm Fe}\st ^{80}{\rm Cr}+ 6n-2e^-+2\nu_e $ & $  0.25 $  &$  7.3  $&$ 31.02  $&$   57.5$  \\
 $  2.135\times 10^{30}$ &  $ 1.197\times 10^{12}$ &   $ ^{80}{\rm Cr}\st ^{74}{\rm Ti}+ 6n-2e^-+2\nu_e $ & $  0.30 $  &$  7.7  $&$ 32.76  $&$   62.1$  \\
 $  2.394\times 10^{30}$ &  $ 1.403\times 10^{12}$ &   $ ^{74}{\rm Ti}\st ^{68}{\rm Ca}+ 6n-2e^-+2\nu_e $ & $  0.36 $  &$  8.1  $&$ 33.57  $&$   67.2$  \\
 $  2.720\times 10^{30}$ &  $ 1.668\times 10^{12}$ &   $ ^{68}{\rm Ca}\st ^{62}{\rm Ar}+ 6n-2e^-+2\nu_e $ & $  0.42 $  &$  8.5  $&$ 34.45  $&$   72.9$  \\
 $  3.145\times 10^{30}$ &  $ 2.016\times 10^{12}$ &   $ ^{62}{\rm Ar}\st ^{56}{\rm  S}+ 6n-2e^-+2\nu_e $ & $  0.47 $  &$  9.0  $&$ 35.44  $&$   79.2$  \\
 $  3.723\times 10^{30}$ &  $ 2.488\times 10^{12}$ &   $ ^{56}{\rm  S}\st ^{50}{\rm Si}+ 6n-2e^-+2\nu_e $ & $  0.53 $  &$  9.4  $&$ 36.57  $&$   86.0$  \\
 $  4.549\times 10^{30}$ &  $ 3.153\times 10^{12}$ &   $ ^{50}{\rm Si}\st ^{42}{\rm Mg}+ 8n-2e^-+2\nu_e $ & $  0.61 $  &$  8.8  $&$ 37.86  $&$   94.5$  \\
\hline
 $  4.624\times 10^{30}$ &  $ 3.472\times 10^{12}$ &   $ ^{42}{\rm Mg}\st ^{36}{\rm Ne}+ 6n-2e^-+2\nu_e $  &&&& \\
&&$^{36}{\rm Ne}+^{36}{\rm Ne}\st ^{72}{\rm Ca}$& $  0.66 $
&$   10.6  $& $ 37.13 $&$    268.2$\\ \hline
 $  5.584\times 10^{30}$ &  $ 4.399\times 10^{12}$ &   $ ^{72}{\rm Ca}\st ^{66}{\rm Ar}+ 6n-2e^-+2\nu_e $ & $  0.69 $  &$  4.8  $&$ 37.56 $&$   31.6$  \\
 $  6.883\times 10^{30}$ &  $ 5.355\times 10^{12}$ &   $ ^{66}{\rm Ar}\st ^{60}{\rm  S}+ 6n-2e^-+2\nu_e $ & $  0.72 $  &$  4.7  $&$ 39.00 $&$   33.5$  \\
 $  8.749\times 10^{30}$ &  $ 6.655\times 10^{12}$ &   $ ^{60}{\rm  S}\st ^{54}{\rm Si}+ 6n-2e^-+2\nu_e $ & $  0.75 $  &$  4.6  $&$ 40.33 $&$   35.2$  \\
\hline
 $  1.157\times 10^{31}$ &  $ 8.487\times 10^{12}$ &   $ ^{54}{\rm Si}\st ^{46}{\rm Mg}+ 8n-2e^-+2\nu_e $ &&&&  \\
&&$^{46}{\rm Mg}+^{46}{\rm Mg}\st ^{92}{\rm Cr}$& $   0.79 $
&$  4.0  $& $ 41.84 $&$    145.3$\\ \hline
 $  1.234\times 10^{31}$ &  $ 9.242\times 10^{12}$ &   $ ^{92}{\rm Cr}\st ^{86}{\rm Ti}+ 6n-2e^-+2\nu_e $ & $  0.80 $  &$  2.0  $&$ 40.88 $&$ 11.4$  \\
 $  1.528\times 10^{31}$ &  $ 1.096\times 10^{13}$ &   $ ^{86}{\rm Ti}\st ^{80}{\rm Ca}+ 6n-2e^-+2\nu_e $ & $  0.82 $  &$  1.9  $&$ 42.04 $&$ 11.4$  \\
 $  1.933\times 10^{31}$ &  $ 1.317\times 10^{13}$ &   $ ^{80}{\rm Ca}\st ^{74}{\rm Ar}+ 6n-2e^-+2\nu_e $ & $  0.83 $  &$  1.8  $&$ 43.31 $&$ 11.2$  \\
 $  2.510\times 10^{31}$ &  $ 1.609\times 10^{13}$ &   $ ^{74}{\rm Ar}\st ^{68}{\rm  S}+ 6n-2e^-+2\nu_e $ & $  0.85 $  &$  1.7  $&$ 44.71 $&$ 10.6$  \\
\hline
 $  3.363\times 10^{31}$ &  $ 2.003\times 10^{13}$ &   $ ^{68}{\rm  S}\st ^{62}{\rm Si}+ 6n-2e^-+2\nu_e $  &&&&\\
&&$^{62}{\rm Si}+^{62}{\rm Si}\st ^{124}{\rm Ni}$&$  0.86 $
&$  1.7  $& $ 46.26 $& $   70.5$ \\ \hline
 $  4.588\times 10^{31}$ &  $ 2.520\times 10^{13}$ &   $ ^{124}{\rm Ni}\st ^{120}{\rm Fe}+ 4n-2e^-+2\nu_e $ & $  0.87 $  &$ 0.8  $&$ 47.78  $&$  3.0$  \\
 $  5.994\times 10^{31}$ &  $ 3.044\times 10^{13}$ &   $ ^{120}{\rm Fe}\st ^{118}{\rm Cr}+ 2n-2e^-+2\nu_e $ & $  0.88 $  &$ 0.9  $&$ 49.65  $&$  2.7$  \\
 $  8.408\times 10^{31}$ &  $ 3.844\times 10^{13}$ &   $ ^{118}{\rm Cr}\st ^{116}{\rm Ti}+ 2n-2e^-+2\nu_e $ & $  0.88 $  &$ 0.8  $&$ 52.28  $&$  2.5$  \\
\hline \hline
\end{tabular}
\end{center}
\end{table*}

\begin{table*}
[t]
\caption{Nuclear processes and released heat in the inner crust, assuming
initial ashes of pure $^{106}{\rm Pd}$ (i.e. as in Table \ref{tab:crust106}) but
suppressing pycnonuclear fusion until $Z=Z_{\rm min}=4$. Only the lines different
than those in the Table \ref{tab:crust106} are presented.
The network  of reactions below pressure
$P=4.624\times 10^{30}$ dyn~cm$^{-2}$ is the same as in the Table  \ref{tab:crust106}.
}
 \label{tab:crust106z}
\begin{center}
\begin{tabular}{llllrrr}
\hline\hline $P_j$ & $\rho_j$ & reactions & $X_n$  & $\Delta
\rho/\rho_j$ & $\mu_{\rm e}$& $Q_j$\\
  (dyn~cm$^{-2}$)  &   (g~cm$^{-3}$)& & &\% & (MeV) & (keV)\\
\hline
 $  4.624\times 10^{30}$ &  $ 3.472\times 10^{12}$ &   $ ^{42}{\rm Mg}\st ^{36}{\rm Ne}+ 6n-2e^-+2\nu_e $ & $  0.66 $  &$  9.9  $&$ 37.13  $&$   76.0$  \\
 $  6.253\times 10^{30}$ &  $ 4.745\times 10^{12}$ &   $ ^{36}{\rm Ne}\st ^{30}{\rm O}+ 6n-2e^-+2\nu_e  $ & $  0.72 $  &$  9.9  $&$ 38.79  $&$   84.2$  \\
 $  9.323\times 10^{30}$ &  $ 6.937\times 10^{12}$ &   $ ^{30}{\rm O}\st ^{24}{\rm C}+ 6n-2e^-+2\nu_e   $ & $  0.77 $  &$  9.4  $&$ 40.89  $&$   91.8$  \\
 $  1.615\times 10^{31}$ &  $ 1.119\times 10^{12}$ &   $ ^{24}{\rm C}\st ^{18}{\rm Be}+ 6n-2e^-+2\nu_e  $ & $  0.84 $  &$  9.2  $&$ 43.59  $&$   94.4$  \\
\hline
 $  3.500\times 10^{31}$ &  $ 2.071\times 10^{13}$ &   $ ^{18}{\rm Be}\st ^{15}{\rm Li}+ 3n-1e^-+1\nu_e $ & && \\
&&$^{15}{\rm Li}+^{15}{\rm Li}\st ^{28}{\rm C}+ 2n$& $  0.88 $
&$   2.3  $&$ 46.78  $&$    152.8$\\ \hline
 $  6.339\times 10^{31}$ &  $ 3.166\times 10^{13}$ &   $ ^{28}{\rm C}\st ^{27}{\rm B}+ 1n-1e^-+1\nu_e $ & $  0.89 $  &$  1.9  $&$  48.98  $&$   8.7$  \\
\hline
\end{tabular}
\end{center}
\end{table*}

\begin{table*}[t]
\caption{
Non-equilibrium processes in the crust of an accreting neutron stars assuming
that the X-ray ashes consist of pure $^{56}{\rm Fe}$.
$P_j$ and $\rho_j$ are pressure and density at which the reaction
takes place.  $\Delta \rho/\rho_j$ is relative density jump
connected with reaction, $Q_j$ is the heat deposited in the
matter.  $X_n$ is the fraction of free neutrons
among  nucleons, and $\mu_e$ is the electron chemical potential,
both in the layer just above the reaction surface.
}
\label{tab:crust56}
\begin{center}
\begin{tabular}{lllllll}
\hline \hline \noalign{\smallskip} $P$ & $\rho$ & Process & $X_n$ &
$\Delta \rho/\rho$  &$\mu_e$ &  $q$
\\ (dyn~cm$^{-2}$) & (g~cm$^{-3}$) & & &  &
&(keV)  \\ \hline \noalign{\smallskip}
$7.23\times 10^{26}$ &
$1.49\times 10^9$ & $^{56}\mathrm{Fe}\rightarrow
^{56}\mathrm{Cr}-2e^-+2\nu_e $ & 0 & 0.08 & 4.08 & 40.7\\
$9.57\times 10^{27}$ & $1.11\times 10^{10}$
&$^{56}\mathrm{Cr}\rightarrow ^{56}\mathrm{Ti}-2e^-+2\nu_e$ & 0 & 0.09 & 8.18& 35.8\\
$1.15\times 10^{29}$ & $7.85\times 10^{10}$
 & $^{56}\mathrm{Ti}\rightarrow ^{56}\mathrm{Ca}-2e^-+2\nu_e$ & 0 & 0.10 & 15.64 & 47.3\\
$4.75\times 10^{29}$ & $2.50\times 10^{11} $
 & $^{56}\mathrm{Ca}\rightarrow
 ^{56}\mathrm{Ar}-2e^-+2\nu_e$ & 0 & 0.11 & 22.48 & 46.1\\
$1.36\times 10^{30}$ & $6.11\times 10^{11} $
 & $^{56}\mathrm{Ar}\rightarrow ^{52}\mathrm{S}+4n-2e^-+2\nu_e$ & 0 & 0.12 & 29.38 & 59.8\\
$1.980\times 10^{30}$  &   $9.075\times 10^{11}$   &
$^{52}\mathrm{S}\rightarrow ^{46}\mathrm{Si}
+6n-2e^-+2\nu_e$   &   0.07   &   0.13&32.27 &128.0\\
$2.253\times 10^{30}$  & $1.131\times 10^{12}$   &   $^{46}\mathrm{Si}\rightarrow
^{40}\mathrm{Mg}+6n-2e^-+2\nu_e$  &  0.18  &0.14  & 32.22 &143.5\\
\hline
$2.637\times 10^{30}$   &   $1.455\times 10^{12}$   &
$^{40}\mathrm{Mg}\rightarrow ^{34}\mathrm{Ne}+6n-2e^-+2\nu_e
$   &  &  && \\
\noalign{\smallskip}
&&
$ ^{34}\mathrm{Ne}+^{34}\mathrm{Ne}\rightarrow ^{68}\mathrm{Ca}$&0.39& 0.17 &34.34&507.9\\
\hline
 $2.771\times 10^{30}$   &   $1.766\times 10^{12}$ &$^{68}\mathrm{Ca}\rightarrow ^{62}\mathrm{Ar}+6n-2e^-+2\nu_e$ &  0.45  &   0.8&34.47&65.8\\
$3.216\times 10^{30}$  &  $2.134\times 10^{12}$  &  $^{62}\mathrm{Ar}\rightarrow
^{56}\mathrm{S}+6n-2e^-+2\nu_e$   &  0.45   &   0.09&35.47 & 71.6  \\
$3.825\times 10^{30}$  &   $2.634\times 10^{12}$  &   $^{56}\mathrm{S}\rightarrow
^{50}\mathrm{Si}+6n-2e^-+2\nu_e$
&   0.50   &   0.09   &  36.59& 77.9\\
$4.699\times 10^{30}$   &  $3.338\times 10^{12}$   &   $^{50}\mathrm{Si}\rightarrow
^{44}\mathrm{Mg}+6n-2e^-+2\nu_e$    &0.55   &0.09  &  37.89& 84.6\\
\hline
\noalign{\smallskip}
$6.043\times 10^{30}$   &   $4.379\times 10^{12}$   &  $^{44}\mathrm{Mg}\rightarrow
^{36}\mathrm{Ne}+8n-2e^-+2\nu_e$  &&&&\\
 &&  $^{36}\mathrm{Ne}+^{36}\mathrm{Ne}\rightarrow ^{72}\mathrm{Ca}$  &&&&\\
 &&$^{72}\mathrm{Ca}\rightarrow
^{66}\mathrm{Ar}+6n-2e^-+2\nu_e$   &   0.61 & 0.14 &39.41 &308.8\\

\hline
\noalign{\smallskip}
$7.233\times 10^{30}$  & $5.839\times 10^{12}$  & $^{66}\mathrm{Ar}\rightarrow
^{60}\mathrm{S}+6n-2e^-+2\nu_e$  &  0.70  &0.04  & 39.01 &29.5\\
$9.238\times 10^{30}$  &  $7.041\times 10^{12}$  &  $^{60}\mathrm{S}\rightarrow
 ^{54}\mathrm{Si}+6n-2e^-+2\nu_e$&
0.73  &  0.04  &40.34 &31.0\\

\hline
\noalign{\smallskip}
$1.228\times 10^{31}$  &  $8.980\times 10^{12}$  &  $^{54}\mathrm{Si}\rightarrow
^{48}\mathrm{Mg}+6n-2e^-+2\nu_e$  &&&& \\
&&$^{48}\mathrm{Mg}+
^{48}\mathrm{Mg}\rightarrow
^{96}\mathrm{Cr}$  &&&&\\
 && $^{96}\mathrm{Cr}\rightarrow
 ^{94}\mathrm{Cr}+2n$ &
0.80  &  0.04 &41.86&135.1\\
\hline
\noalign{\smallskip}
$1.463\times 10^{31}$  & $1.057\times 10^{13}$  & $^{94}\mathrm{Cr}\rightarrow
^{88}\mathrm{Ti}+6n-2e^-+2\nu_e$  &  0.81  &0.02  & 41.99 &11.5\\

$1.816\times 10^{31}$  & $1.254\times 10^{13}$  & $^{88}\mathrm{Ti}\rightarrow
^{82}\mathrm{Ca}+6n-2e^-+2\nu_e$  &  0.82  &0.02  & 43.18 &11.3\\

$2.304\times 10^{31}$  & $1.506\times 10^{13}$  & $^{82}\mathrm{Ca}\rightarrow
^{76}\mathrm{Ar}+6n-2e^-+2\nu_e$  &  0.84  &0.02  & 44.48 &10.9\\

$2.998\times 10^{31}$  & $1.838\times 10^{13}$  & $^{76}\mathrm{Ar}\rightarrow
^{70}\mathrm{S}+6n-2e^-+2\nu_e$  &  0.85  &0.02  & 45.91 &10.0\\
\hline
\noalign{\smallskip}
$4.028\times 10^{31}$  &  $2.287\times 10^{13}$  &  $^{70}\mathrm{S}\rightarrow
^{64}\mathrm{Si}+6n-2e^-+2\nu_e$  &&&& \\
&&$^{64}\mathrm{Si}+
^{64}\mathrm{Si}\rightarrow
^{128}\mathrm{Ni}$  &&&&\\
 && $^{128}\mathrm{Ni}\rightarrow
 ^{126}\mathrm{Ni}+2n$ &
0.87  &  0.01 &47.48&67.3\\
\hline
\noalign{\smallskip}

$5.278\times 10^{31}$  & $2.784\times 10^{13}$  & $^{126}\mathrm{Ni}\rightarrow
^{124}\mathrm{Fe}+2n-2e^-+2\nu_e$  &  0.88  &0.01  & 48.50 &2.5\\

$7.311\times 10^{31}$  & $3.493\times 10^{13}$  & $^{124}\mathrm{Fe}\rightarrow
^{122}\mathrm{Cr}+2n-2e^-+2\nu_e$  &  0.89  &0.01  & 51.05 &2.4\\

\hline
\hline
\end{tabular}
\end{center}
\end{table*}

\begin{table*}
[t]
\caption{Nuclear processes and released heat in the inner crust, assuming
initial ashes of pure $^{56}{\rm Fe}$ (i.e. as in Table \ref{tab:crust56}) but
suppressing pycnonuclear fusion until $Z=Z_{\rm min}=4$. Only the lines different
than those in the Table \ref{tab:crust56} are presented.
The network  of reactions below pressure
$P=2.637\times 10^{30}$ dyn~cm$^{-2}$ is the same as
in the Table  \ref{tab:crust56}.
}
 \label{tab:crust56z}
\begin{center}
\begin{tabular}{llllrrr}
\hline\hline $P$ & $\rho$ & reactions & $X_n$  & $\Delta
\rho/\rho$ & $\mu_{\rm e}$& $q$\\
  (dyn~cm$^{-2}$)  &   (g~cm$^{-3}$)& & &\% & (MeV) & (keV)\\
\hline
 $  2.637\times 10^{30}$ &  $ 1.455\times 10^{12}$ &   $ ^{40}{\rm Mg}\st ^{34}{\rm Ne}+ 6n-2e^-+2\nu_e $ & $  0.40 $  &$  16.0 $&$ 34.45  $&$   161.4$  \\
 $  3.227\times 10^{30}$ &  $ 1.961\times 10^{12}$ &   $ ^{34}{\rm Ne}\st ^{28}{\rm  O}+ 6n-2e^-+2\nu_e $ & $  0.50 $  &$  17.9 $&$ 35.71  $&$   183.4$  \\
 $  4.254\times 10^{30}$ &  $ 2.831\times 10^{12}$ &   $ ^{28}{\rm O}\st ^{22}{\rm C}+ 6n-2e^-+2\nu_e $ & $  0.61 $  &$  19.4  $&$ 37.49  $&$   209.8$  \\
 $  6.392\times 10^{30}$ &  $ 4.541\times 10^{12}$ &   $ ^{22}{\rm C}\st ^{16}{\rm Be}+ 6n-2e^-+2\nu_e  $ & $  0.72 $  &$  20.0  $&$ 39.89  $&$   238.2$  \\
\hline
 $  1.216\times 10^{31}$  &  $8.617\times 10^{12}$  &  $^{16}\mathrm{Be}\rightarrow^{13}\mathrm{Li}+4n-1e^-+1\nu_e$  &&&& \\
&&$^{13}\mathrm{Li}+^{13}\mathrm{Li}\rightarrow ^{26}\mathrm{C}$  &&&&\\
 && $^{26}\mathrm{C}\rightarrow ^{24}\mathrm{C}+2n$ & 0.79  &  7.5 &43.18 & 331.2\\
\hline
\noalign{\smallskip}
 $  1.727\times 10^{31}$ &  $ 1.186\times 10^{13}$ &   $ ^{24}{\rm C}\st ^{18}{\rm Be}+ 6n-2e^-+2\nu_e  $ & $  0.85 $  &$  7.4  $&$ 43.64  $&$   84.1$  \\
\hline
 $  3.798\times 10^{31}$  &  $2.200\times 10^{12}$  &  $^{18}\mathrm{Be}\rightarrow^{17}\mathrm{Li}+1n-1e^-+1\nu_e$  &&&& \\
&&$^{17}\mathrm{Li}+^{17}\mathrm{Li}\rightarrow ^{34}\mathrm{C}$  &&&&\\
 && $^{34}\mathrm{C}\rightarrow ^{30}\mathrm{C}+4n$ & 0.88  &  0.028 &46.86 & 142.0\\
\hline
\noalign{\smallskip}
\hline
\end{tabular}
\end{center}
\end{table*}

\end{appendix}
\end{document}